\documentclass{article}

\usepackage{arxiv}

\usepackage[utf8]{inputenc} 
\usepackage[T1]{fontenc}    
\usepackage{hyperref}       
\usepackage{url}            
\usepackage{booktabs}       
\usepackage{amsfonts}       
\usepackage{nicefrac}       
\usepackage{microtype}      
\usepackage{lipsum}
\usepackage{graphicx}

\usepackage{array}
\usepackage{stfloats}
\usepackage{verbatim}
\usepackage{cite}
\usepackage{amsmath,amssymb,amsfonts}
\usepackage{algorithmic}
\usepackage{textcomp}
\usepackage{xcolor}
\usepackage{algorithm}
\usepackage{tcolorbox}
\usepackage{diagbox}
\usepackage{mathtools}
\usepackage{multirow}
\usepackage{url}
\usepackage{makecell}
\usepackage{booktabs}
\usepackage{caption}
\usepackage{subcaption}
\usepackage{tikz}

\graphicspath{ {./images/} }

\title{MAP-UOT: A Memory-Efficient Approach to Unbalanced Optimal Transport Implementation}

\author{
 Chengyu Sun \\
  \texttt{sunchengyu@hnu.edu.cn} \\
   \And
 Jinyu Hu \\
  College of Computer Science and Electronic Engineering\\
  Hunan University\\
  \texttt{hujinyu@hnu.edu.cn} \\
  \And
 Hong Jiang \\
  Computer Science and Engineering Department\\
  University of Texas at Arlington\\
  \texttt{hong.jiang@uta.edu} \\
}

\begin{document}
\maketitle
\begin{abstract}
Unbalanced optimal transport (UOT) has been widely used as a fundamental tool in many application domains, where it often dominates the application running time.
While many researchers have proposed various optimizations for UOT, few have attempted to optimize it from a computer architecture's perspective.
In this paper, we first study the performance bottlenecks of UOT through a series of experiments, which reveals that UOT is heavily memory-bound. Guided by these findings, we propose MAP-UOT, a \underline{M}emory-efficient \underline{AP}proach to the implementation and optimization of UOT on CPU and GPU platforms.
Our experimental evaluations show that the proposed strategy consistently and significantly outperforms the state-of-the-art (SOTA) implementations. Specifically, it provides single-threaded performance improvement over POT/COFFEE by up to 2.9X/2.4X, with an average of 1.9X/1.6X.
At the same time, it provides parallelized performance improvement over POT/COFFEE by up to 2.4X/1.9X, with an average of 2.2X/1.8X, on Intel Core i9-12900K; and over POT by up to 3.5X, with an average of 1.6X, on Nvidia GeForce RTX 3090 Ti.
MAP-UOT also shows great performance improvement on the Tianhe-1 supercomputer.
\end{abstract}


\section{Introduction}
\label{sec:introduction}

The unbalanced optimal transport (UOT) problem with the entropic regularization scheme is the extension and continuation of optimal transport (OT) problems~\cite{lee2019parallel}.
The focus of research on UOT has been to solve a linear optimization problem, that is, minimize the overhead of transporting mass between two histograms.
In recent years, UOT has been continuously developed into an increasingly important tool employed in a wide range of application domains utilizing machine learning (ML)~\cite{pham2020unbalanced,chizat2018unbalanced}, such as
computational biology~\cite{schiebinger2019optimal}, computational imaging~\cite{lee2019parallel}, neuroimaging~\cite{gramfort2015fast}, natural language processing~\cite{kusner2015word}, domain adaptation~\cite{ferradans2014regularized,courty2014domain,flamary2016optimal}, supervised learning~\cite{frogner2015learning}, etc.

However, it is noted that its computational complexity of $O(N^2)$ remains a performance barrier to these applications.
After experimental and theory analysis, we found that the execution time consumed by UOT is more than half of the total in all four representative applications, and this dominance in total execution time by UOT increases with the matrix size.
As a result, many implementations of UOT problem have been proposed to decrease its costly computational overhead and increase its scalability.
The existing studies mainly focus on analyzing the convergence characteristics of UOT for faster convergence speed~\cite{genevay2016stochastic}, proposing specific solutions for specific data structures~\cite{sato2020fast}, and using ML models to reformulate UOT to obtain results through training~\cite{yang2018scalable}.
The most famous and most widely deployed algorithm among these algorithms is the Sinkhorn algorithm~\cite{knight2008sinkhorn}, a solver with entropy regularization optimizations.
The Sinkhorn algorithm is more attractive than other optimization strategies because its implementation is easy and practical, which only alternately rescales all rows and all columns of the given matrix, and its application to UOT achieves almost optimal convergence~\cite{genevay2016stochastic}.
As a result, we choose UOT with the Sinkhorn algorithm solver (hereinafter referred to as the UOT algorithm) as baseline to conduct our research.

Previous works, e.g., COFFEE~\cite{sun2023coffee} has optimized the algorithm on CPU clusters using MPI.
However, it fails to recognize the memory-bound problem of the algorithm and fails to propose new strategies to address this problem.
It also cannot take advantage of other computing hardware such as GPUs.
Moreover, it is challenging to pinpoint with analysis alone where the performance bottleneck of the algorithm lies and then design an effective solution.
In multi/many-cores CPU implementation, the challenge involves taking advantage of different levels of optimization to achieve optimal performance while fully exploiting parallelism.
In GPU implementation, the challenge is designing data and memory layout, along with an appropriate tiling strategy, to effectively manage the more sensitive performance variations.

By establishing a global memory Roofline model~\cite{williams2009roofline} and conducting a theoretical analysis (Section~\ref{sec:motivation}), we conclude that the UOT algorithm is heavily memory-bound.
Therefore, it is essential for a memory-bound algorithm to improve memory access efficiency and increase computational intensity.
With this insight, we propose MAP-UOT, a \underline{M}emory-efficient \underline{AP}proach that optimizes UOT according to its memory-bound characteristics.
The main design of MAP-UOT is interweaving the row rescaling and column rescaling that are originally performed independently, resulting in substantially reduction of the memory access.
The key to this novel interweaving strategy, which is in spirit similar that of blocked dense matrix multiplication, is to embed in each row (column) rescaling the partial column (row) rescaling of the column (row) whose element is participating in the current row (column) rescaling.
In other words, this approach combines the row and column rescaling in a double-loop of each iteration, which enables the matrix to be read and written only once, as opposed to reading and writing the matrix twice by the state-of-the-art (SOTA) implementations with separate row rescaling and column rescaling.
Moreover, MAP-UOT focuses on reducing the noncontinuous memory access of column rescaling to make the approach cache-friendly.
For the GPU architecture, we further optimize the most time-consuming parts by designing memory-efficient data layout and tiling strategies, which is combined with the memory hierarchy features of prefetching and vectorization to achieve high performance.

It is noted that our strategy has broad generality for algorithms with similar iterations of row and column rescaling.
It can be applicable to higher dimensional tensors as well.
For example, the Earth Mover's Distance (EMD) algorithm~\cite{cuturi2013sinkhorn} can also use our strategy to reduce memory access.

The main contributions of this paper are as follows:
\begin{itemize}
  \item We identify potential performance bottlenecks of the UOT algorithm through a series of experiments. To the best of our knowledge, this is the first time that a performance analysis for identifying UOT performance bottlenecks has been proposed.
  \item We propose a memory-efficient strategy from the perspective of computer architecture based on the identified performance bottlenecks. We implement and optimize the algorithm on advanced multi/many-cores CPU and GPU architectures and supercomputer.
  \item We conduct experiments that demonstrate the significant performance improvements over existing implementations. Our design exhibits wide applicability and scalability across multiple platforms.
\end{itemize} 

\section{Background and Related Work}
\label{sec:background}

\subsection{Unbalanced Optimal Transport}

\begin{figure}[t]
\centering
{\includegraphics[width=0.95\linewidth]{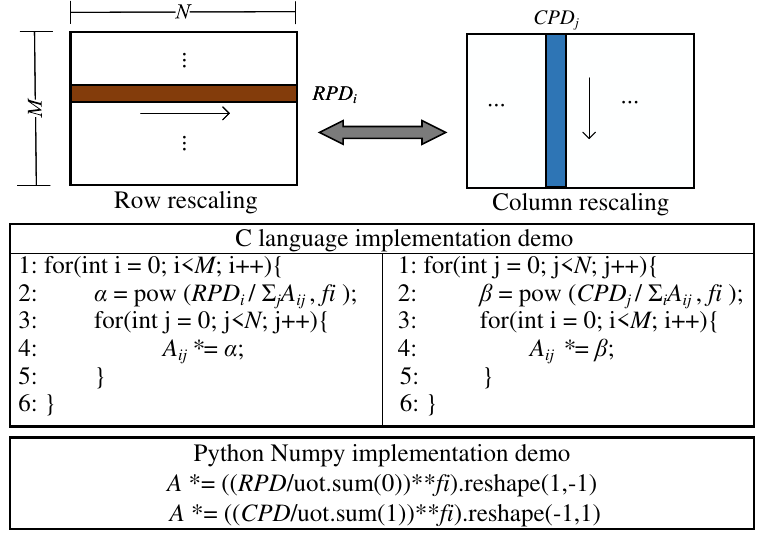}}
\caption{Implementation of the UOT algorithm with C language and Python language implementation demos.}
\label{fig:typical}
\end{figure}

The baseline implementation of the UOT algorithm consists of row rescaling and column rescaling as shown in Figure~\ref{fig:typical} with C language and Python language implementations.
The figure illustrates an example of a matrix $A = \left(A_{ij}\right)$ with $M$ rows and $N$ columns, where $fi = er/(er+ep)$, $er$ and $ep$ are the coefficients related to UOT applications.
Row probability distribution ($RPD$) and column probability distribution ($CPD$) are vector metrics of length $M$ and $N$, respectively, which are the constraints of the algorithm.
$\alpha$ ($\beta$) is a rescaling factor for the row (column) rescaling.
The row rescaling factor of the $i^{th}$ row is calculated as $\alpha=pow(RPD_{i}/\sum_{j=0}^{N-1} A_{ij}, fi)$, and the same calculation applies to the column rescaling factor.
Performing a row rescaling of the matrix means multiplying each data element in the matrix by the rescaling factor of the corresponding row, and the same process again goes for column rescaling.
The main idea of the UOT algorithm is to iterate between row rescaling and column rescaling alternately until a set error limit is reached.
It should be noted that performing row rescaling first and then column rescaling or vice versa does not affect the theoretical limit value, and generally does not make a distinction in practical applications~\cite{knight2008sinkhorn}.

\subsection{Proportion of Application Time Consumed by the UOT Algorithm}

\begin{figure}[t]
\centering
{\includegraphics[width=1.0\linewidth]{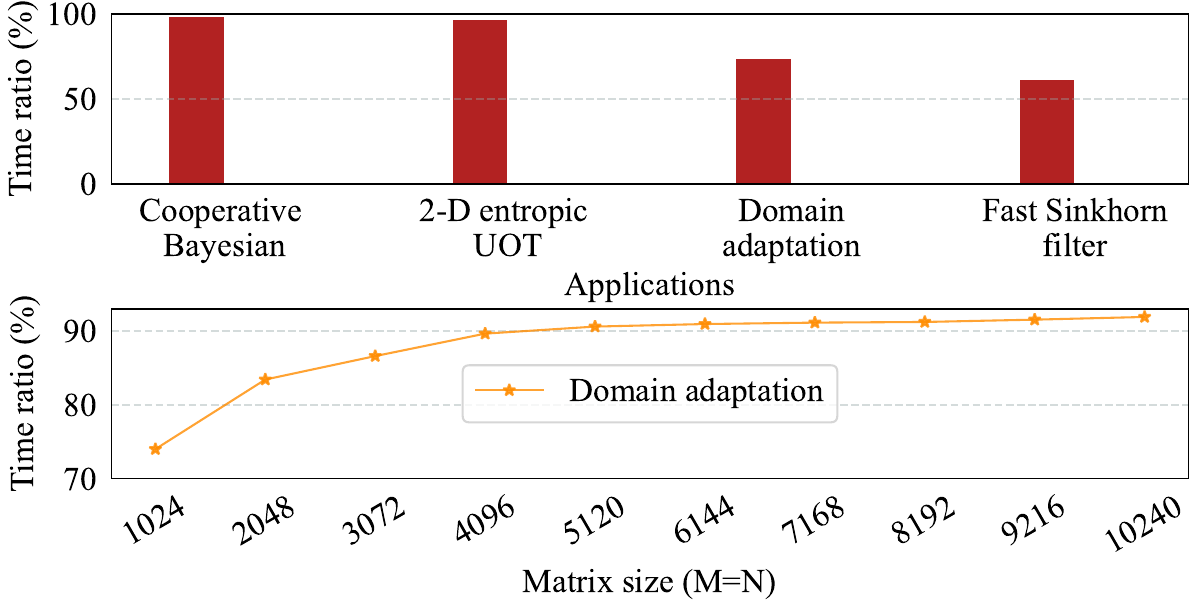}}
\caption{The proportion of time occupied by the UOT algorithm among four applications (top)~\cite{sun2023coffee} and the proportion of time occupied by the UOT algorithm in the domain adaptation application (bottom).}
\label{fig:Time_ratio}
\end{figure}

Previous studies have observed that the UOT algorithm consistently dominates the end-to-end execution times of applications~\cite{wang2020sequential,flamary2016optimal,pham2020unbalanced,pai2021fast}.
To better understand UOT's impact on the application execution time, COFFEE~\cite{sun2023coffee} experimentally examined four applications, including cooperative Bayesian~\cite{wang2020sequential}, 2-D entropic UOT~\cite{flamary2021pot,pham2020unbalanced}, domain adaptation~\cite{flamary2021pot,flamary2016optimal} and fast Sinkhorn filter~\cite{pai2021fast}, that adopt the UOT algorithm.
As shown in Figure~\ref{fig:Time_ratio}, the execution time of UOT occupies 99\%, 97\%, 74\% and 62\% of the end-to-end execution times of the four applications respectively with a matrix size of $M = N = 1024$.
Moreover, we select the domain adaptation application~\cite{flamary2016optimal} to evaluate the impact of matrix size on UOT's time dominance.
It shows that the proportion of end-to-end application time spent by the UOT algorithm increases with the matrix size, because of its computational complexity of $O(N^2)$.
Therefore, optimizing this critical part will play a critical role in improving the performance of the entire application.

\subsection{Related Work}

Most of the existing UOT implementations aim to speed up the rate of convergence.
Lee et al.~\cite{lee2019parallel} developed a new UOT model based on the Beckmann formulation, making it possible to significantly reduce the number of variables.
Genevay et al.~\cite{genevay2016stochastic} proposed a stochastic optimization algorithm based on three possible scenarios (discrete OT, semi-discrete OT, and continuous OT) to acquire faster convergence properties.
Sato et al.~\cite{sato2020fast} proposed a method that can solve UOT problems with tree metrics in one-dimensional (1-D) space in quasi-linear time. Bonneel et al.~\cite{bonneel2019spot} proposed a method to quickly process UOT problem of different cardinality point sets supporting constant distributions by 1-D slicing.
Furthermore, some researchers try to use the idea of ML to optimize UOT problem.
Yang et al.~\cite{yang2018scalable} proposed a scalable UOT solution based on generative adversarial network. Based on this, transport maps can be learned by generative models when a transport cost is not available.
Arjovsky et al.~\cite{arjovsky2017wasserstein,gulrajani2017improved} made a good reformulation based on the 1-Wasserstein distance, but the use of the approximate 1-Lipschitz functions creates difficulties in practical applications.
It is noted that many researchers have pointed out that UOT problem is easy to be parallelized, and some of them have proposed parallel algorithms~\cite{lee2019parallel,genevay2016stochastic,altschuler2019massively,chizat2018scaling}.
However, little research is known to actually implement it on parallel software or hardware architectures (e.g., Message Passing Interface (MPI), GPU, FPGA).
It is noted that a recent work in 2021~\cite{tithi2021new} from the Intel parallel computing team have proposed a different design for Intel's new PIUMA multi-core architecture, and implemented it in a shared memory environment to accelerate the Sinkhorn-Knopp (SK) algorithm in parallel. Moreover, on the basis of the work of Intel, a work in 2023~\cite{sun2023coffee} called COFFEE has proposed a cross-layer design for fast and efficient executions of the SK algorithm on HPC systems.
The authors of COFFEE have designed CPU-oriented and MPI-oriented optimizations to speed up the algorithm on multi-cores CPUs and supercomputers.
Compared with theirs, our design proposes a novel memory-efficient approach based on the discovered memory-bound characteristics of UOT and demonstrates wider applicability and scalability. 

\section{Identifying Performance Bottlenecks of UOT}
\label{sec:motivation}

In this section, we evaluate the UOT performance based on the two representative libraries Numpy and Cupy on multi/many-cores CPUs and GPUs, respectively, in order to identify performance bottlenecks.
It is noted that both implementations are from the Python Optimal Transport (POT) library~\cite{flamary2021pot}, an open-source Python library that provides SOTA solvers for optimal transport-related algorithms.
In the rest of the paper, the matrices are assumed to be stored in the row-major format without loss of generality.

\subsection{Performance Bottleneck}

\begin{figure}[t]
\centering
{\includegraphics[width=1.0\linewidth]{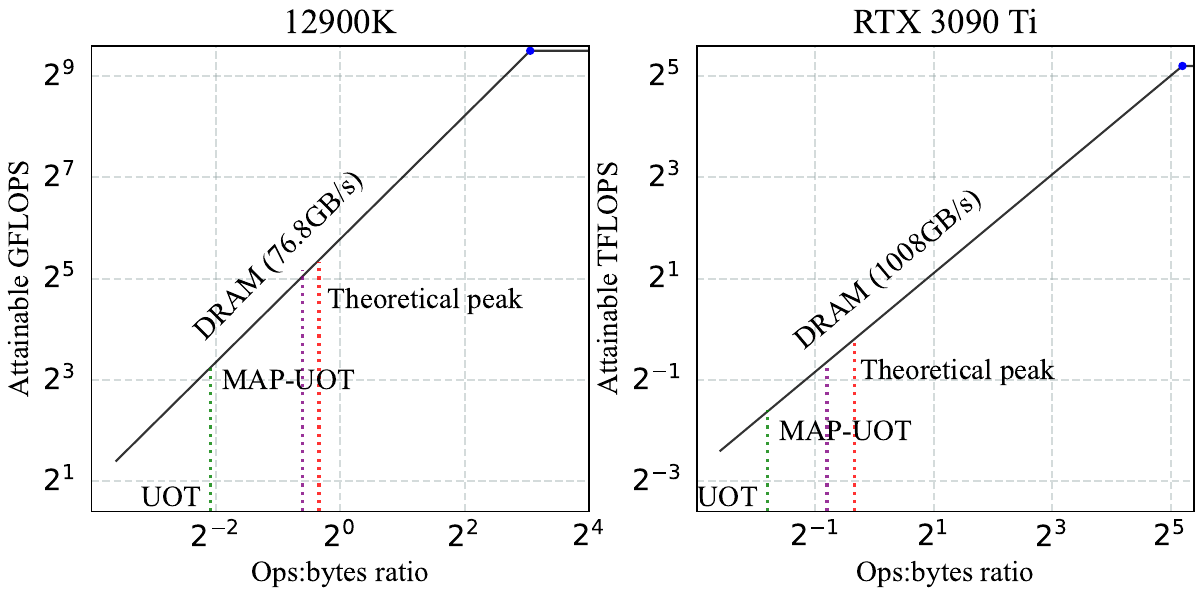}}
\caption{Global memory Roofline model of UOT on 12900K and RTX 3090 Ti, respectively.}
\label{fig:Roofline}
\end{figure}

To investigate the root causes of the above observed time dominance of the UOT algorithm in applications and thus identify its potential performance bottlenecks, we plot the global memory Roofline models~\cite{williams2009roofline,ding2019instruction} for the 12900K CPU and RTX 3090 Ti GPU architectures.
The baseline implementation in Figure~\ref{fig:typical} shows that the matrix operations consist of mainly row rescaling and column rescaling.
Taking row rescaling as an example (6 lines of pseudo-code), there are $M$ ADD operations of $i$ for Line 1, $M \times N$ LOAD and ADD operations of $A_{ij}$, $M$ DIV operations, and $M$ power function calls for Line 2, $M \times N$ ADD operations of $j$ for Line 3, and $M \times N$ LOAD, MUL and STORE operations of $A_{ij}$ for Line 4.
Thus, the memory traffic $Q$ is $3 \times M \times N$, and the work $W$ is $3 \times M \times N + 3 \times M$.
For the sake of simplicity, in the Roofline model, ADD, DIV, MUL and CALL are all counted as one calculation.
The term operational intensity is $I = W / Q$, i.e., operations per byte of DRAM traffic.
So the operational intensity of the UOT algorithm (assuming FP32) is:
\begin{equation}\label{1}
  I = \frac{6 \times M \times N + 6 ( M + N )}{6 \times M \times N \times sizeof(Type)} = \frac{M \times N + M + N}{4 \times M \times N}
\end{equation}
As shown in Figure~\ref{fig:Roofline}, $I$ is approximately equal to $1 / 4$ (green dotted lines), which is extremely low because the peak performance can theoretically be exploited only when the operational intensity $I$ reaches 10.3 and 39.7 on 12900K and RTX 3090 Ti, respectively.
It is noted that the two values are the inflection points (blue points) of the two figures respectively on the two curves in the Roofline model, obtained by dividing the peak performance by the peak bandwidth.
Therefore, the UOT algorithm is heavily memory-bound.
Moreover, based on the Roofline model, the theoretical minimum $Q$ value is $2 \times M \times N$, assuming that the cache can fully accommodate matrix $A$. Because at least one read and one write of the entire matrix $A$ are required.
Therefore, as shown in Figure~\ref{fig:Roofline}, the theoretical limit that the UOT algorithm can reach are the theoretical peak lines (red dotted lines).
We also plot purple dotted lines showing the performance attainable by MAP-UOT. It is found that the performance achieved by MAP-UOT is close to the theoretical limit.

\begin{figure}[t]
\centering
{\includegraphics[width=1.0\linewidth]{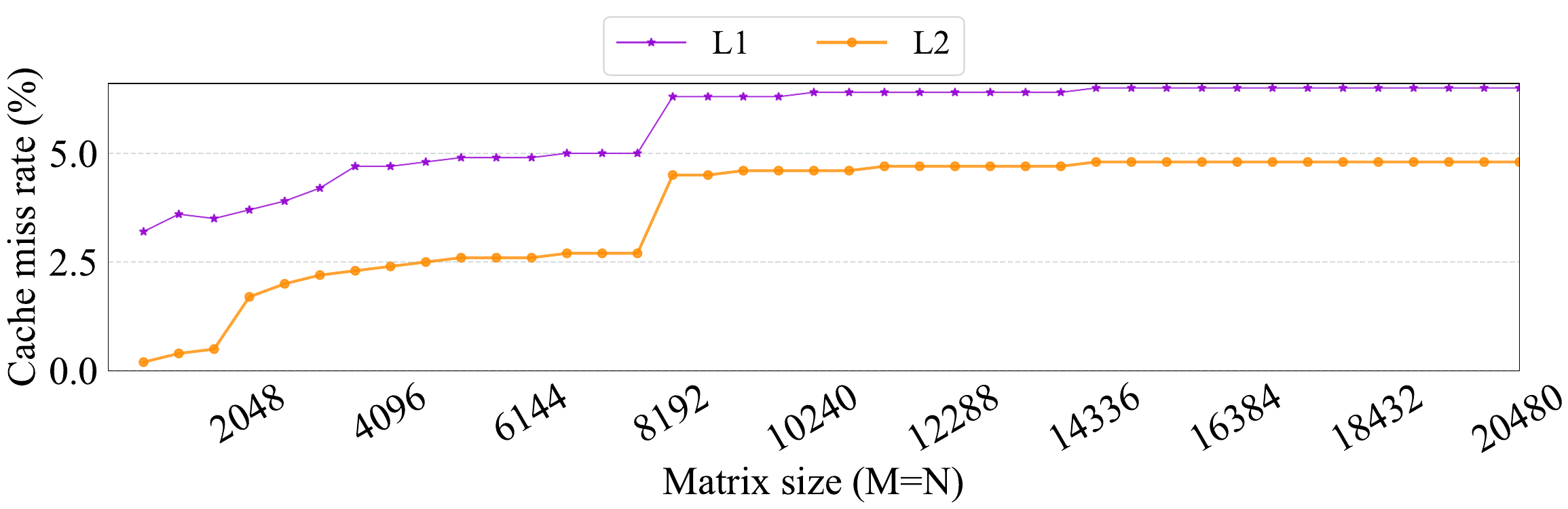}}
\caption{L1 and L2 cache miss rate of UOT on 12900K with Numpy implementation.}
\label{fig:Motivation_cache}
\end{figure}

\begin{figure}[t]
\centering
{\includegraphics[width=1.0\linewidth]{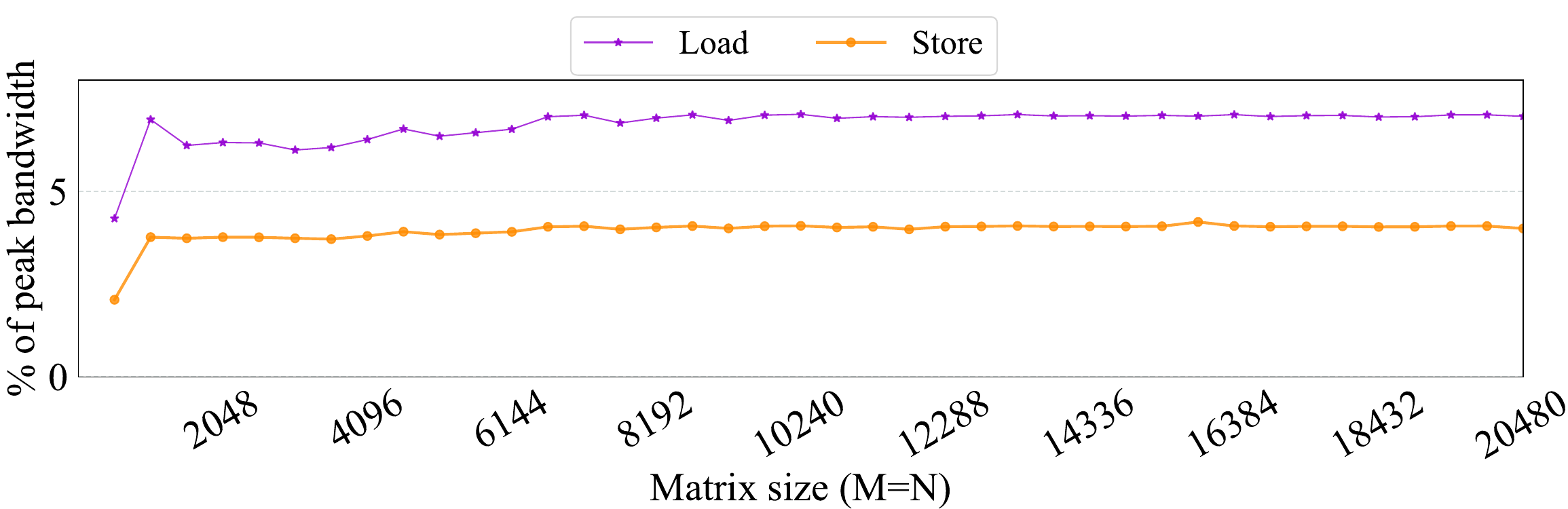}}
\caption{Global load/store throughput of UOT on RTX 3090 Ti with Cupy implementation.}
\label{fig:Motivation_throughput}
\end{figure}

Then, we implement different experiments on the two architectures for further exploration.
It is found that the memory accesses of the column rescaling of the matrix $A$ in the baseline implementation are cache-unfriendly, because referencing different cache lines when sequentially accessing $A[i][j]$ and $A[i + 1][j]$ makes it difficult to take advantage of cache locality.
As shown in Figure~\ref{fig:Motivation_cache}, on 12900K, for example, when the matrix size is $10240 \times 10240$, L1 cache miss rate and L2 cache miss rate are $6.4\%$ and $4.6\%$, respectively.
As a comparison, the L1 cache miss rate of dense matrix multiplication in OpenBLAS can reach $3.2\%$ in similar matrix size~\cite{wang2015design}.
Therefore, there is a potential to reduce the cache misses using more efficient design.
For GPUs, we measure global load/store throughput when implementing the UOT algorithm with Cupy.
The results are shown in Figure~\ref{fig:Motivation_throughput}.
It is observed that global throughput is only a small fraction of the peak bandwidth.
For example, when the matrix size is $10240 \times 10240$, global load throughput can only reach about $7.07\%$ of the peak, and global store throughput can only reach about $4.07\%$ of the peak.
This is also because memory accesses of the column rescaling are cache-unfriendly.

\subsection{Key Insight for MAP-UOT Design}

The above important findings and insights motivate us to explore efficient ways to minimize memory accesses so that the performance is close to the theoretical limit of the Roofline model for the UOT algorithm.
Thus, the key idea of MAP-UOT is to reduce memory accesses using novel interweaving design, by doing partial row (column) rescaling while carrying out column (row) rescaling to increase cache locality, and maximize the operational intensity, along with a series of related optimizations.
For the implementation on multi/many-cores CPU, MAP-UOT ensures that each iteration is done using only one double-loop, and that memory accesses are cache-friendly.
For the implementation on GPU, MAP-UOT combines different features of the GPU architecture to further optimize the algorithm.
We first redesign the UOT algorithm according to the same high-level idea and find the most time-consuming parts.
Then we further optimize memory accesses by designing workload and memory allocation strategies, and adding vectorization instructions and prefetching technique.

\section{Design and Implementation of MAP-UOT}
\label{sec:implementation detail}

In this section, we introduce the design and implementation of MAP-UOT on multi/many-cores CPU and GPU architectures.
To achieve efficient memory access, we need to increasing the usage of matrix data in cache before replacing it.
To this end, we carry out in-depth analysis of the characteristics of the UOT algorithm and find ways to interweave row rescaling and column rescaling.
More specifically, the implementation for each iteration of MAP-UOT is divided into four parts.
That is, calculate the row rescaling factor based on the sum of rows of the matrix (part \textcircled{\raisebox{-0.9pt}{\textbf{1}}}); modify the matrix according to the row rescaling factor and accumulate the sum of columns of the matrix at the same time (part \textcircled{\raisebox{-0.9pt}{\textbf{2}}}); calculate the column rescaling factor based on the sum of columns of the matrix (part \textcircled{\raisebox{-0.9pt}{\textbf{3}}}); modify the matrix according to the column rescaling factor and calculate the sum of rows of the matrix at the same time (part \textcircled{\raisebox{-0.9pt}{\textbf{4}}}).
It is discovered that part \textcircled{\raisebox{-0.9pt}{\textbf{2}}} and part \textcircled{\raisebox{-0.9pt}{\textbf{4}}} dominate the running time of whole process because of their computational complexity of $O(N^2)$, while the computational complexity of the other parts are $O(N)$, so they are the key to optimization and also the focus of the rest of this section.

\subsection{Design and Optimization on Multi/many-cores CPU}

\subsubsection{Design Methodology}
\begin{figure}[t]
\centering
{\includegraphics[width=1.0\linewidth]{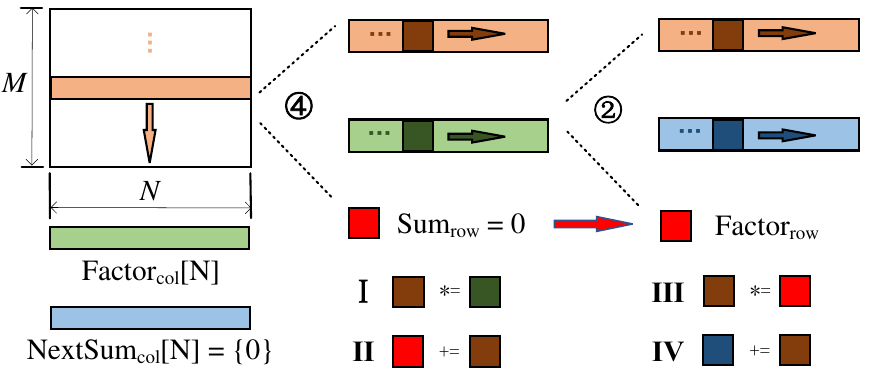}}
\caption{Schematic of MAP-UOT on multi/many-cores CPU.}
\label{fig:CPU}
\end{figure}

\begin{algorithm}
\caption{Algorithm of MAP-UOT on multi/many-cores CPU.}
\begin{algorithmic}[1]
\label{alg:Pthreads_algorithm}
\REQUIRE{$A[M\times N]$, $Factor_{col}[N]$, $NextSum_{col}[T\times N]$ initialized as 0s, $CPD[N]$, $RPD[M]$, $Sum_{row}$ initialized as 0, $fi$, $T$ (number of threads).}
\ENSURE{$A[M\times N]$ after once row and column rescaling.}
\FOR{$j$ from 0 to \emph{N-1}}
\STATE{$Factor_{col}[j] = pow(CPD[j]/Factor_{col}[j], fi)$;}
\ENDFOR
\STATE{//Double-loop for every thread with id $tid$}
\FOR{$i$ from $tid*M/T$ to \emph{(tid+1)*M/T-1}}
\FOR{$j$ from 0 to \emph{N-1}}
\STATE{$A[i][j] *= Factor_{col}[j]$;}
\STATE{$Sum_{row} += A[i][j]$;}
\ENDFOR
\STATE{$Sum_{row} = pow(RPD[i]/Sum_{row}, fi)$;}
\FOR{$j$ from 0 to \emph{N-1}}
\STATE{$A[i][j] *= Sum_{row}$;}
\STATE{$NextSum_{col}[tid][j] += A[i][j]$;}
\ENDFOR
\ENDFOR
\FOR{$i$ from 0 to \emph{T-1}}
\FOR{$j$ from 0 to \emph{N-1}}
\STATE{$Factor_{col}[j] += NextSum_{col}[i][j]$;}
\ENDFOR
\ENDFOR
\end{algorithmic}
\end{algorithm}

We propose the design of MAP-UOT as shown in Figure~\ref{fig:CPU} on multi/many-cores CPU.
For simplicity, we only give the implementation of first column rescaling and then row rescaling.
While our design and optimizations are equally applicable to the case of first row rescaling and then column rescaling.

For the initialization of the algorithm, an array $Factor_{col}$ of length $N$ is preprocessed to store column rescaling factor.
It is calculated as $Factor_{col}[j] = pow(CPD[j]/\sum_{i=0}^{M-1} A_{ij}, fi)$.
We also need an array $NextSum_{col}$ of length $N$ to store the sum of column for next iteration of column rescaling, and a variable $Sum_{row}$ for current row rescaling.
Then we introduce the main double-loop of our design.
Almost all computational tasks are done within this double-loop.
The double-loop is performed in row order.
We divide the traversal of a row in the inner loop into two steps in Figure~\ref{fig:CPU}, corresponding to part \textcircled{\raisebox{-0.9pt}{\textbf{4}}} and part \textcircled{\raisebox{-0.9pt}{\textbf{2}}} described above.
In part \textcircled{\raisebox{-0.9pt}{\textbf{4}}}, each piece of data in a row is accessed in turn and used to implement two computations.
Computation \textbf{I}: multiply each data by the column rescaling factor at the corresponding position of the $Factor_{col}$ array.
This completes the column rescaling for the current row.
The next work for this row that needs to be done is row rescaling.
Computation \textbf{II}: add each data that has been column rescaling to $Sum_{row}$.
When finishing the traversal of this row, the sum of this row is accumulated in $Sum_{row}$.
Then, $Sum_{row}$ is calculated with the corresponding $RPD$ and $fi$ to get the $Factor_{row}$ used for the row rescaling of this row.
In part \textcircled{\raisebox{-0.9pt}{\textbf{2}}}, we need to access each data of the row and perform two computations again, which is similar to part \textcircled{\raisebox{-0.9pt}{\textbf{4}}}.
Computation \textbf{III}: multiply each data by the $Factor_{row}$ obtained in the previous step.
The row rescaling of this row is completed.
Computation \textbf{IV}: add each data that has been row rescaling to the corresponding position of the $NextSum_{col}$ array.
So far all operations on this row have been completed.
Each next row only needs to repeat the above process.
After completing this double-loop, the entire matrix has completed one column rescaling and one row rescaling at a time.
Finally, the $NextSum_{col}$ array will accumulate the sum of columns of the matrix, it can be calculated with $CPD$ and $fi$ to get the column rescaling factors that need to be used in the next iteration.

MAP-UOT can bring a significant performance improvement over existing algorithms.
Because when operating on one row, it can complete both row rescaling and column rescaling for this row.
In this way, the utilization of each data can be maximized.
Only one double-loop is needed to complete each iteration.
In Numpy implementation, four double-loops are required, that is, calculating the sum of columns of the matrix, modifying the matrix according to the column rescaling factor, calculating the sum of rows of the matrix, and modifying the matrix according to the row rescaling factor.
As a result, MAP-UOT reduces memory access instructions and thus improves operational intensity.
As long as the cache can accommodate the row currently being operated on, MAP-UOT can approach the theoretical peak of the UOT algorithm.
Another point is that MAP-UOT has completely continuous access to data, which is cache-friendly and helps to achieve high performance.

\subsubsection{Parallel Implementation}

In this subsection, we perform further parallel optimizations on the basis of the above design.
Because the parallel programming model Pthreads can more intuitively display the data allocation and calculation of each thread, we take Pthreads instead of OpenMP as an example to demonstrate the optimization of MAP-UOT on multi/many-cores CPU.
The matrix is first divided into several submatrices by rows to the individual threads, which makes the most sense since all computations are done in row order.
That is, each of the $T$ threads handles $M/T$ rows of the matrix.
It is found that MAP-UOT is very easy to parallelize, each thread will do the exact same work, as shown in Algorithm~\ref{alg:Pthreads_algorithm}, Line 5-15.
$NextSum_{col}[tid][]$ ($tid$ refers to the id of the thread, $0<=tid<T$) corresponds to $NextSum_{col}$ array in Figure~\ref{fig:CPU}, which refers to the sum of columns of the submatrix.
All threads will be joined into the main thread when they have finished their work.
Then main thread can calculate the sum of columns of the matrix, which is obtained by adding all $NextSum_{col}[tid][]$ to the $Factor_{col}$ array, as shown in Algorithm~\ref{alg:Pthreads_algorithm}, Line 16-20.
The whole process is highly parallel, the computational complexity $O(N)$ work of main thread and the overhead of thread launching and joining can be negligible.
To further improve performance, we unroll the loop to reduce loop overhead.
Then we add data-level parallelism on the basis of the above optimization.
We choose the AVX2 series instruction to optimize Line 7-8, 12-13 in Algorithm~\ref{alg:Pthreads_algorithm}.

\subsection{Design and Optimization on GPU}

\subsubsection{Design Methodology}

In this subsection we will introduce the design of MAP-UOT on GPU, which is very different from optimization on multi/many-cores CPU in detail.
Despite the fact that we will adopt similar high-level ideas, it is still a challenge to take full advantage of the architectural features of GPU to achieve high performance.
For example, adopting a similar design of multi/many-cores CPU, that is, to have each thread responsible for the operations on one or several rows of the matrix in order to complete one iteration in a double-loop, is unreasonable and inefficient.
Because this cannot take advantage of the characteristics of GPU at all.
The number of threads is too small or even less than the number of CUDA cores, not enough for GPU scheduling.
Moreover, when summing the columns of a matrix, a large number of threads will access the same position of the $Sum_{col}$ array at the same time, resulting in extremely low performance.
Therefore, we need an elaborate design for part \textcircled{\raisebox{-0.9pt}{\textbf{2}}} and part \textcircled{\raisebox{-0.9pt}{\textbf{4}}} which is completely different from the design on multi/many-cores CPU.

\subsubsection{Implementation Details of Part \textcircled{\raisebox{-0.9pt}{\textbf{2}}}}

\begin{figure}[t]
\centering
{\includegraphics[width=1.0\linewidth]{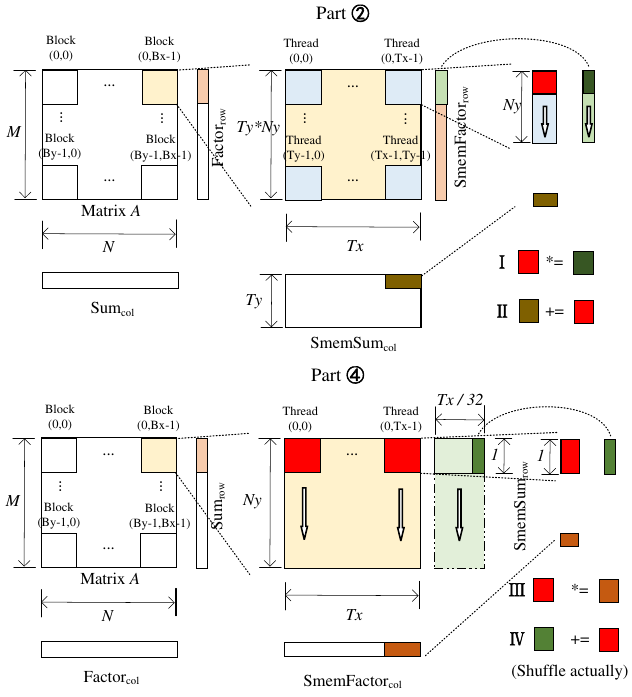}}
\caption{Schematic of MAP-UOT on GPU.}
\label{fig:GPU}
\end{figure}

The upper part of Figure~\ref{fig:GPU} shows our design of MAP-UOT specifically for part \textcircled{\raisebox{-0.9pt}{\textbf{2}}}.
First we introduce the tiling strategy.
We have arranged a two-dimensional (2-D) grid and 2-D blocks for this kernel.
The grid consists of $By \times Bx$ blocks, and each block consists of $Ty \times Tx$ threads.
Our design ensures that all warp accesses are coalesced so that the maximum number of memory requests can be satisfied with the fewest memory events, since coalescing memory access is extremely important for a memory-intensive kernel~\cite{cheng2014professional}.

Then we introduce the process of part \textcircled{\raisebox{-0.9pt}{\textbf{2}}} and memory arrangement.
Each block will load a submatrix of ($Ty \times Ny) \times Tx$, and the block needs to first modify the submatrix with the row rescaling factor and then calculate the sum of columns of the submatrix and add it to the $Sum_{col}$ array.
It is necessary to use shared memory (Smem) to improve the performance.
We use $SmemFactor_{row}$ to load the $Factor_{row}$ array needed for this block, because the $Factor_{row}$ array needs to be accessed multiple times.
Then we use $SmemSum_{col}$ of $Ty \times Tx$ to get the sum of columns of the submatrix.
More specifically for a thread, let it be responsible for the processing of $Ny$ data in the submatrix.
Each thread will first multiply the data of a row by the corresponding data in $SmemFactor_{row}$.
Then it will add the data to the corresponding position of $SmemSum_{col}$, for example, add the data processed by thread (i, j) to the $SmemSum_{col}[i][j]$.
Each thread will repeat this process $Ny$ times, as shown in Algorithm~\ref{alg:step_2}, Line 6-10.
It is noted that the reason for having one thread process $Ny$ data instead of processing one data is to help hide memory access latency.
Moreover, we use 128-bit vectorization load/store instructions to further improve memory access.
We also use the idea of loop unrolling and preload the register with the data for the next row of the submatrix in each round of the loop to further hide memory access latency.
Next, we need to further process the data obtained in $SmemSum_{col}$, that is, perform a reduction operation on it to calculate the sum of columns of data in $Ny$ rows and store it in the first row.
Finally, the results of the sum of columns of the submatrix need to be added to the $Sum_{col}$ array. At this time, many blocks will access an address at the same time, causing the results to be wrong.
We use atomic operation instructions to ensure access security and get correct results, as shown in Algorithm~\ref{alg:step_2}, Line 12-15.

\begin{center}
\begin{minipage}{0.95\linewidth}
\begin{algorithm}[H]
\caption{Algorithm of part \textcircled{\raisebox{-0.9pt}{\textbf{2}}} on GPU.}
\begin{algorithmic}[1]
\label{alg:step_2}
\STATE{$\_\_shared\_\_ $ $ SmemFactor_{row}[Ty \times Ny];$}
\STATE{$\_\_shared\_\_ $ $ SmemSum_{col}[Ty][Tx];$}
\STATE{$idx_r = Ny \times (By \times blockIdx.y + threadIdx.y);$}
\STATE{$idx_c = Bx \times blockIdx.x + threadIdx.x;$}
\STATE{$idx = idx_r \times N + idx_c;$}
\FOR{$i$ from 0 to \emph{Ny-1}}
\STATE{$A[idx] *= SmemFactor_{row}[Ny \times threadIdx.y + i];$}
\STATE{$SmemSum_{col}[threadIdx.y][threadIdx.x]+=A[idx];$}
\STATE{$idx += N;$}
\ENDFOR
\STATE{$\_\_syncthreads();$}
\STATE{$Reduce(SmemSum_{col}[0][threadIdx.x]$ $\leftarrow$ \par $\quad SmemSum_{col}[][threadIdx.x]);$}
\IF{$threadIdx.y==0$}
\STATE{$atomicAdd(\&Sum_{col}[idx_c],$ \par $\quad SmemSum_{col}[0][threadIdx.x]);$}
\ENDIF
\end{algorithmic}
\end{algorithm}
\end{minipage}
\end{center}

\begin{algorithm}
\caption{Algorithm of part \textcircled{\raisebox{-0.9pt}{\textbf{4}}} on GPU.}
\begin{algorithmic}[1]
\label{alg:step_4}
\STATE{$\_\_shared\_\_ $ $ SmemFactor_{col}[Tx];$}
\STATE{$\_\_shared\_\_ $ $ SmemSum_{row}[Tx >> 5];$}
\STATE{$idx_r = Ny \times blockIdx.y;$}
\STATE{$idx_c = Bx \times blockIdx.x + threadIdx.x;$}
\STATE{$idx = idx_r \times N + idx_c;$}
\STATE{$laneID = threadIdx.x \& 31;$}
\STATE{$warpID = threadIdx.x >> 5;$}
\FOR{$i$ from 0 to \emph{Ny-1}}
\STATE{$A[idx] *= SmemFactor_{col}[threadIdx.x];$}
\STATE{$unsigned$ $int$ $u = 16,t=A[idx];$}
\WHILE{$u >>= 1$}
\STATE{$t += \_\_shfl\_down(t, u);$}
\ENDWHILE
\IF{$laneID==0$}
\STATE{$SmemSum_{row}[warpID] = t;$}
\ENDIF
\STATE{$Reduce(SmemSum_{row}[0]$ $\leftarrow$ $SmemSum_{row}[]);$}
\IF{$threadIdx.x==0$}
\STATE{$atomicAdd(\&Sum_{row}[idx_r],SmemSum_{row}[0]);$}
\ENDIF
\STATE{$\_\_syncthreads();$}
\STATE{$idx_r++;$}
\STATE{$idx += N;$}
\ENDFOR
\end{algorithmic}
\end{algorithm}

\subsubsection{Implementation Details of Part \textcircled{\raisebox{-0.9pt}{\textbf{4}}}}

The design of part \textcircled{\raisebox{-0.9pt}{\textbf{4}}} and part \textcircled{\raisebox{-0.9pt}{\textbf{2}}} have similarities, we will focus on the differences.
In terms of tiling strategy, we have arranged 1-D blocks and each block consists of $Tx$ threads.
Then for each block, we use $SmemFactor_{col}$ to load the corresponding position of the $Factor_{col}$ array, and use the $SmemSum_{row}$ array to store the sum of the currently processed row from the submatrix.
Each thread will first multiply the data of a row by the corresponding data of $SmemFactor_{col}$, and then cooperate with other threads to obtain the sum of the currently processed row of the submatrix and finally add it to the corresponding position of the $Sum_{row}$ array.
Each thread will repeat this process $Ny$ times to complete this part.

In the process of getting the sum of the currently processed row, it is necessary to perform a reduction operation instead of adding it directly to the $SmemSum_{row}$ array.
We achieve this using the warp shuffle instruction.
The shuffle instruction can make the threads in the same warp exchange data directly without using Smem and have lower latency than Smem.
As shown in Algorithm~\ref{alg:step_4}, Line 11-17, the reduction operation is divided into two steps.
First, we use the shuffle instruction to obtain the result of a warp and store it in $SmemSum_{row}$, and then perform a reduction operation again to obtain the reduction result of $SmemSum_{row}$.
Finally, we use the atomic instruction to store the reduction result in the $Sum_{row}$ array of global memory, as shown in Algorithm~\ref{alg:step_4}, Line 18-20. 

\section{Experimental Evaluation}
\label{sec:evaluation}

\subsection{Experimental Setup}

In this section, we evaluate MAP-UOT over POT~\cite{flamary2021pot} and COFFEE~\cite{sun2023coffee} to demonstrate the performance improvement.
The experimental environment includes multi-cores CPU, GPU and supercomputer.
Different parameters of the advanced architectures are listed in Table~\ref{hardware}.
Our code is written in C language and is packaged as a library function that can be called by Python.
All the code was compiled with -O3 flag.
For the sake of simplicity, we use FP32 to demonstrate the experimental results.
It is noted that we obtain similar performance improvement when using double-precision floating-point numbers.

\begin{figure}[t]
\centering
{\includegraphics[width=1.0\linewidth]{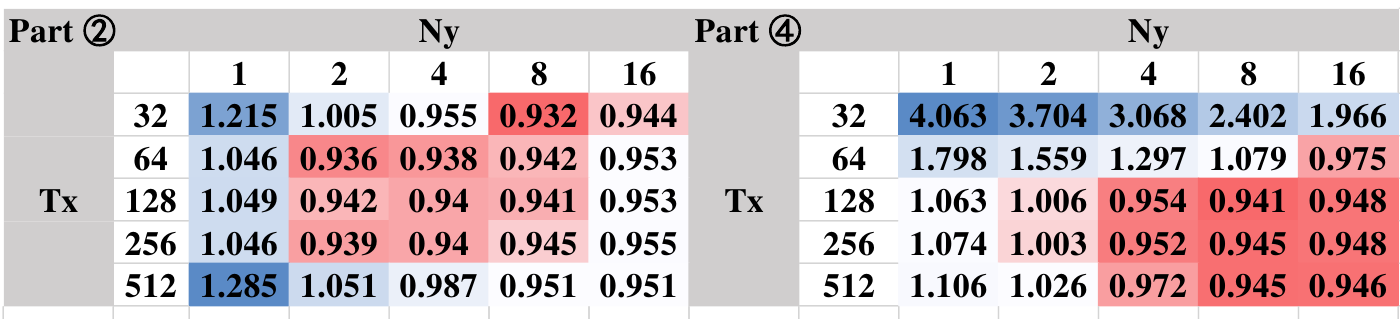}}
\caption{Running time (ms) of MAP-UOT over different parameters on RTX 3090 Ti.}
\label{fig:Para_time}
\end{figure}

We first show the performance improvements of MAP-UOT on CPU in serial environment and on GPUs.
Then, we focus on evaluating the improvement of different metrics on the two platforms.
On 12900K, we measure the scalability of multiple threads, the cache misses reduction, and the performance impact of Pthreads false sharing issues.
On RTX 3090 Ti, we measure throughput improvements and memory consumption reductions.
Moreover, we demonstrate the scalability of MAP-UOT at large scale on Tianhe-1 supercomputer~\cite{yang2010th}.
At last, we present the results of the overall improvement in the end-to-end application.
In experiments related to GPU, the choice of GPU parameters plays a crucial role in performance improvement, especially in the choice of the tiling parameters.
So we first experimentally select the value that yields optimal performance.
We set $Ty=2$ to show the performance improvement of different $Tx$, $Ny$ on a $10240 \times 10240$ matrix.
As shown in Figure~\ref{fig:Para_time}, it is discovered that $Tx=32$, $Ny=8$ in part \textcircled{\raisebox{-0.9pt}{\textbf{2}}}, and $Tx=128$, $Ny=8$ in part \textcircled{\raisebox{-0.9pt}{\textbf{4}}} could achieve the best performance.

\begin{table}[!h]
    \centering
    \caption{Hardware evaluation platform.}
    \scalebox{0.7}{
    \begin{tabular}{llll}
    \toprule
        \multicolumn{4}{c}{12th Gen Intel Core i9-12900K}                                    \\
    \midrule
        Number of Cores         & 16                   & Max Frequency    & 5.1/3.9 GHz      \\
        Peak Performance        & 793.6 GFLOPS (FP32)  & Memory Bandwidth & 76.8 GB/s        \\
        OS kernel               & Linux 5.15           & Compiler         & GCC 9.4.0        \\
    \toprule
        \multicolumn{4}{c}{NVIDIA GeForce RTX 3090 Ti}                                       \\
    \midrule
        GPU Boost Clock         & 1.86 GHz             & Peak Performance & 40 TFLOPS (FP32) \\
        Memory                  & 384-bit GDDR6X       & Memory Capacity  & 24 GB            \\
        Memory Clock            & 1.313 GHz            & Memory Bandwidth & 1008 GB/s        \\
        CUDA Cores              & 10752                & CUDA Version     & 11.4             \\
        NVIDIA GPU Driver       & 515.65               & Compiler         & NVCC V11.4.48    \\
    \toprule
        \multicolumn{4}{c}{Tianhe-1 supercomputer~\cite{yang2010th}}                         \\
        CPU                  & Intel Xeon Westmere & Cores per Node           & 12           \\
        Max Frequency        & 2.93 GHz            & Memory Capacity          & 32 GB        \\
        Interconnect Network & Infiniband QDR      & Network Bandwidth        & 160 Gb/s     \\
    \midrule
    \bottomrule
    \end{tabular}}
    \label{hardware}
\end{table}

\begin{figure}[t]
\centering
{\includegraphics[width=1.0\linewidth]{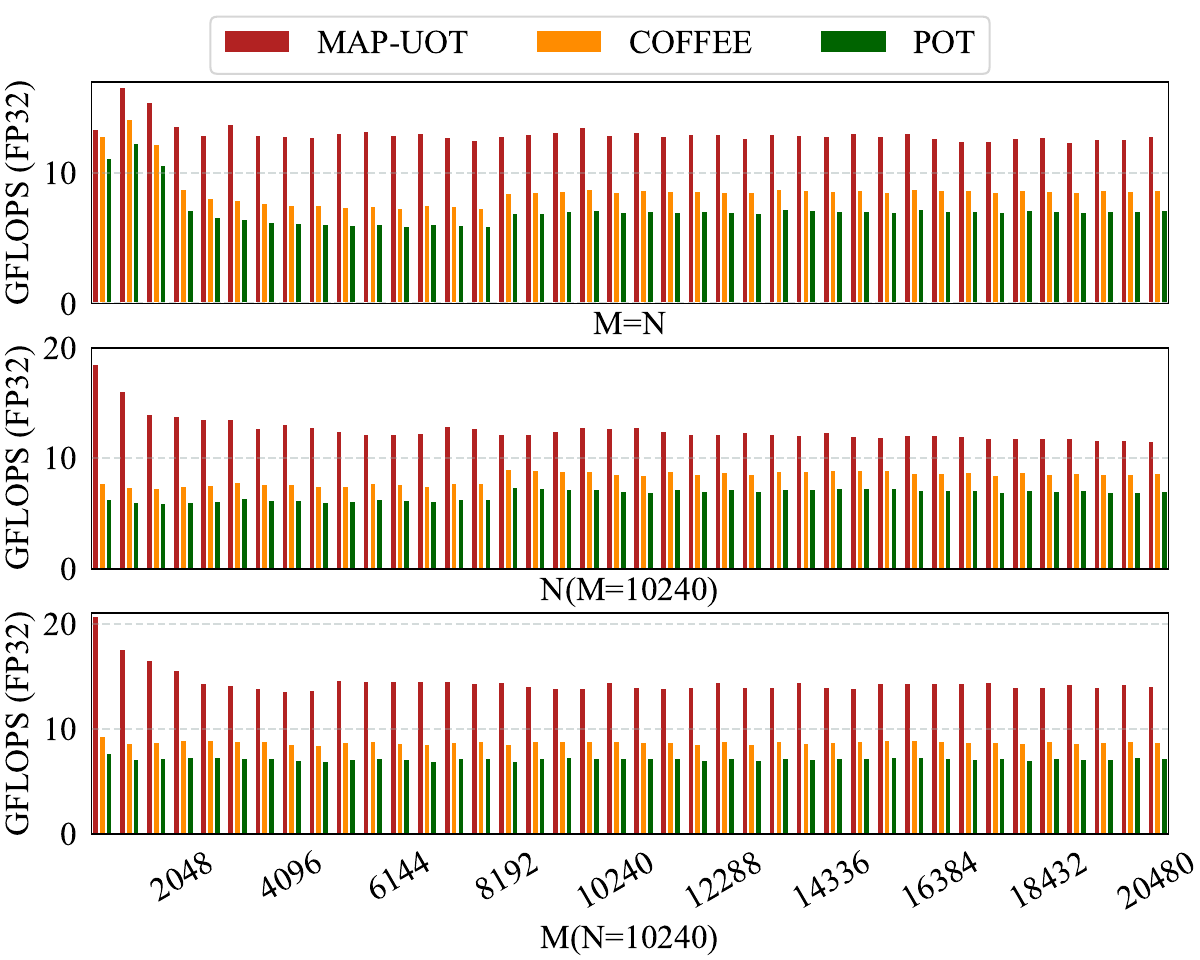}}
\caption{Single-threaded performance of UOT on 12900K.}
\label{fig:CPU_new}
\end{figure}

\begin{figure}[t]
\centering
{\includegraphics[width=1.0\linewidth]{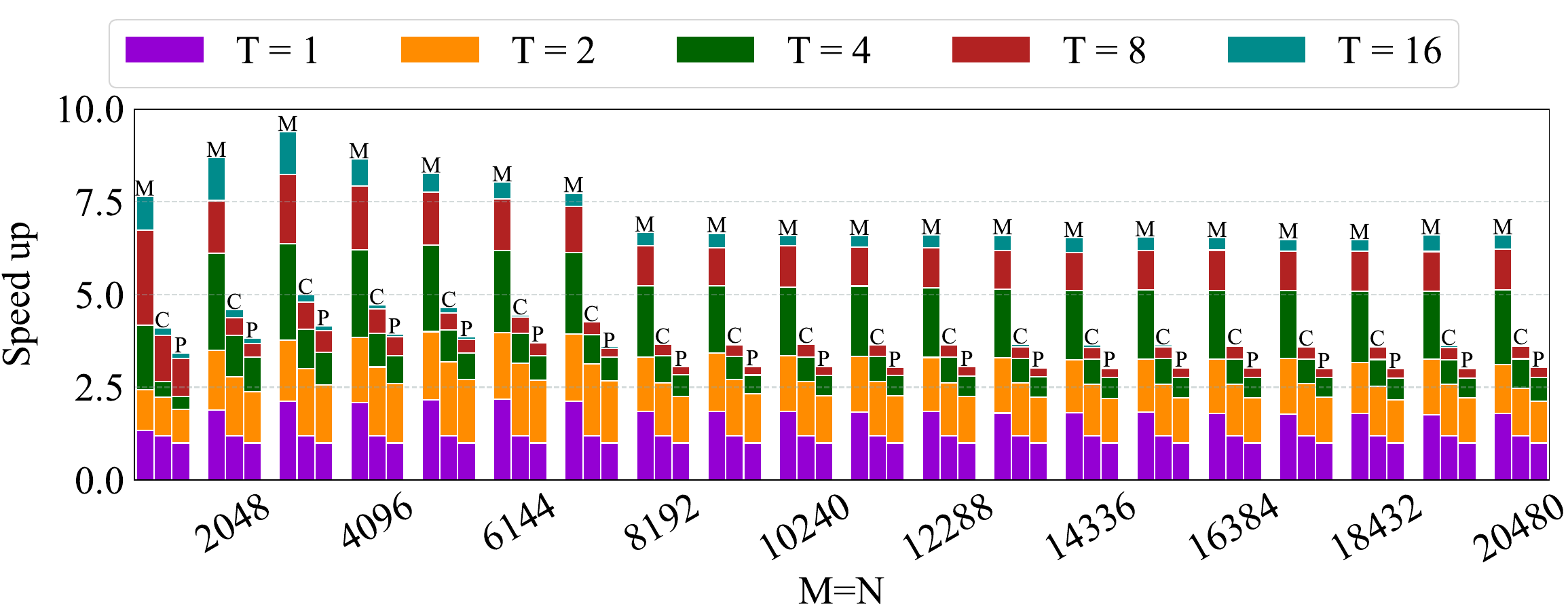}}
\caption{Scalability of multiple threads for UOT. The legend 'T' refers to the number of threads, 'M' refers to MAP-UOT, 'P' refers to POT and 'C' refers to COFFEE.}
\label{fig:CPU_scalable}
\end{figure}

\begin{figure}[t]
\centering
{\includegraphics[width=1.0\linewidth]{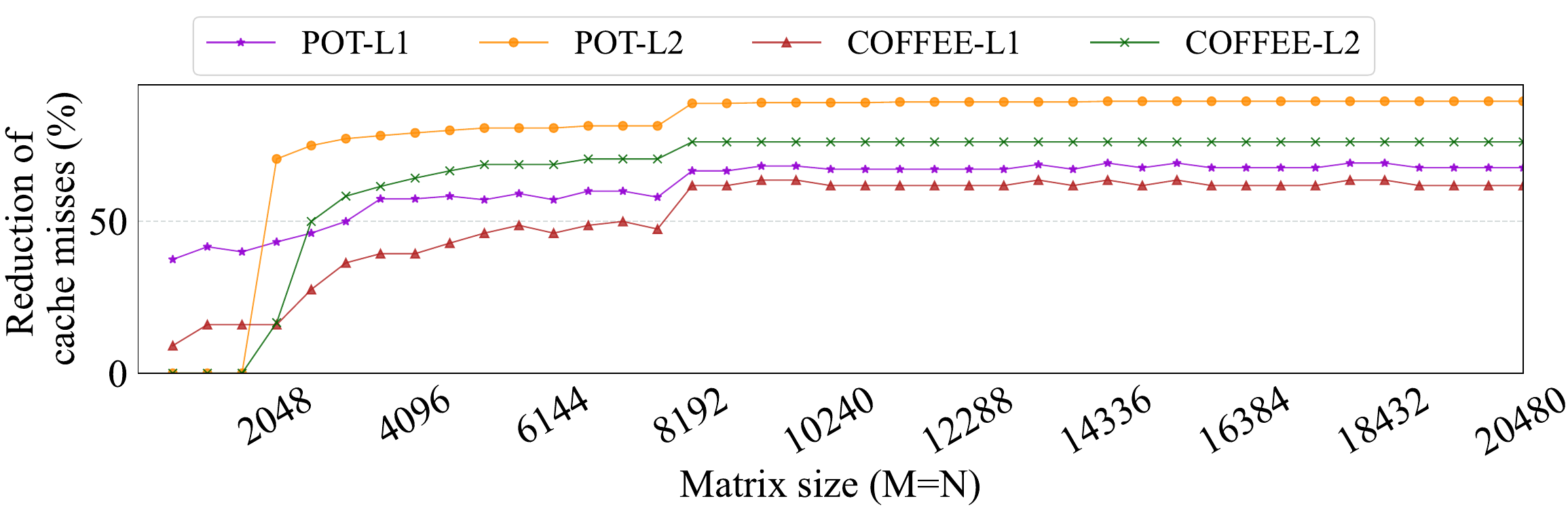}}
\caption{Reduction of cache misses over POT and COFFEE for UOT.}
\label{fig:CPU_cache}
\end{figure}

\begin{figure}[t]
\centering
{\includegraphics[width=1.0\linewidth]{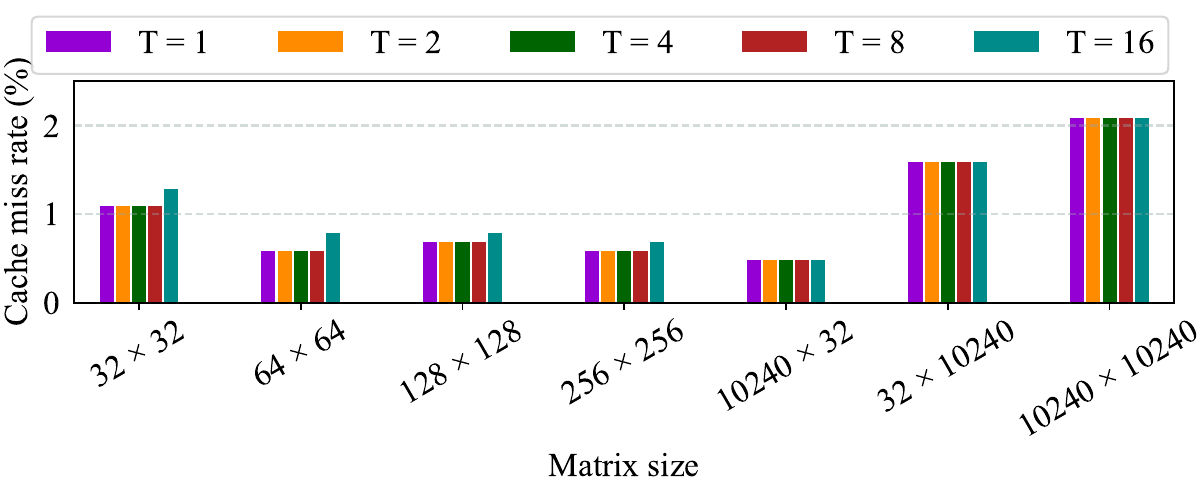}}
\caption{L1 cache miss rate on different number of threads of MAP-UOT. The legend 'T' refers to the number of threads.}
\label{fig:CPU_pthreads}
\end{figure}

\begin{figure}[t]
\centering
{\includegraphics[width=1.0\linewidth]{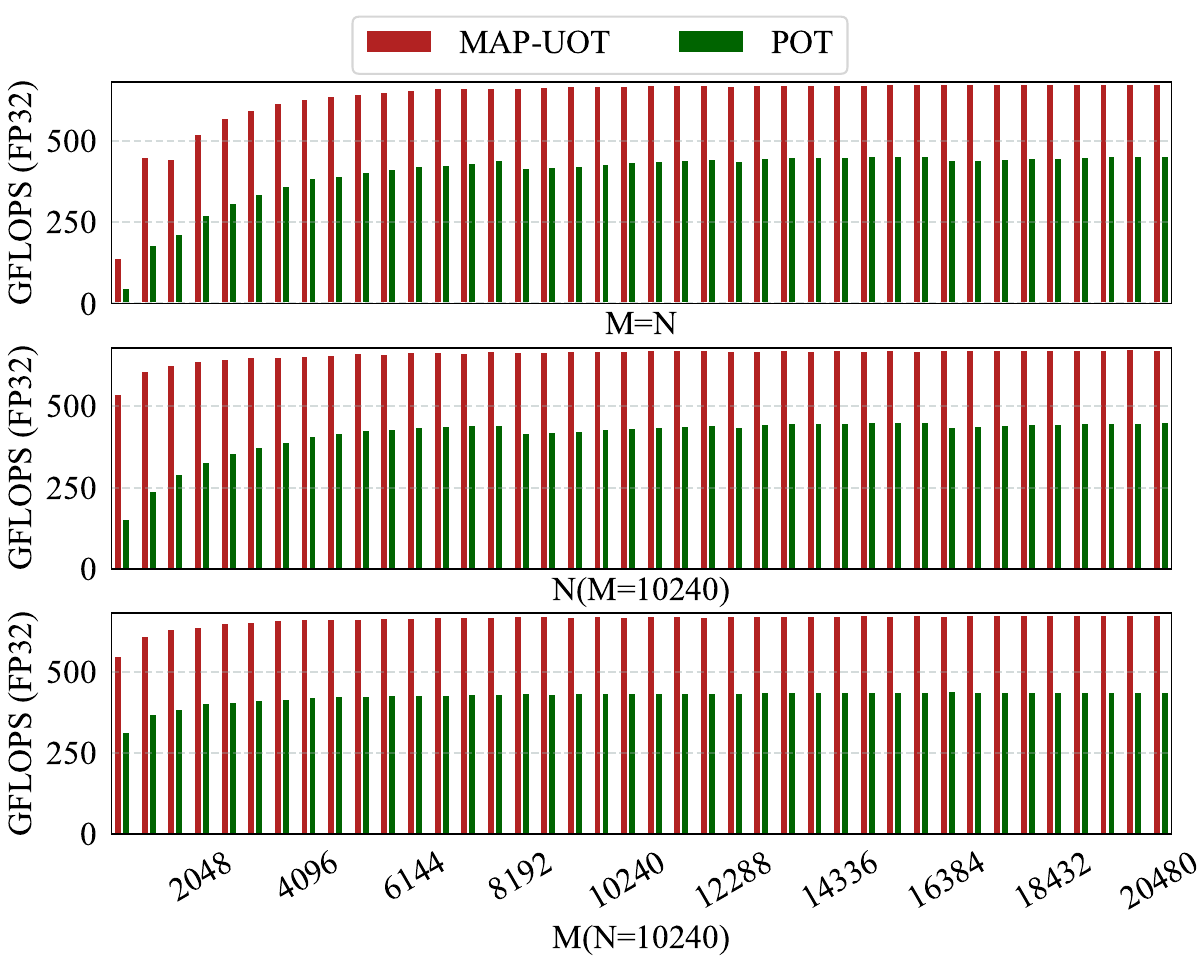}}
\caption{Performance of UOT on RTX 3090 Ti.}
\label{fig:GPU_new}
\end{figure}

\begin{figure}[t]
\centering
{\includegraphics[width=1.0\linewidth]{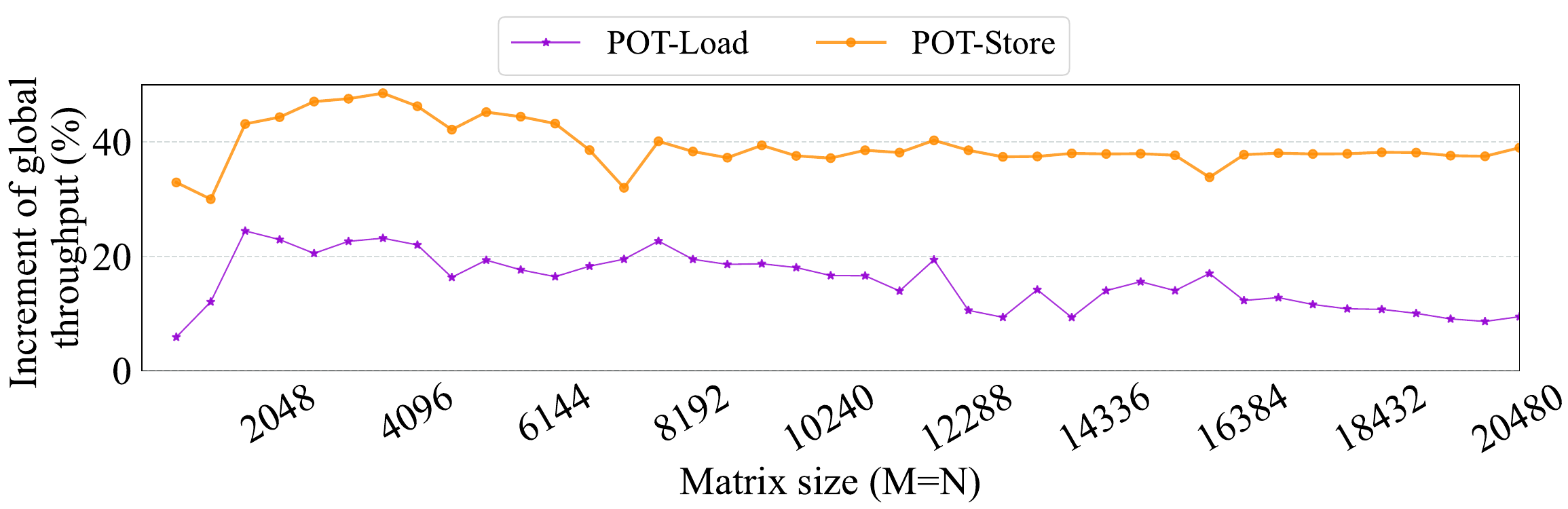}}
\caption{Increment of global throughput over POT for UOT.}
\label{fig:GPU_throughput}
\end{figure}

\begin{figure}[t]
\centering
{\includegraphics[width=1.0\linewidth]{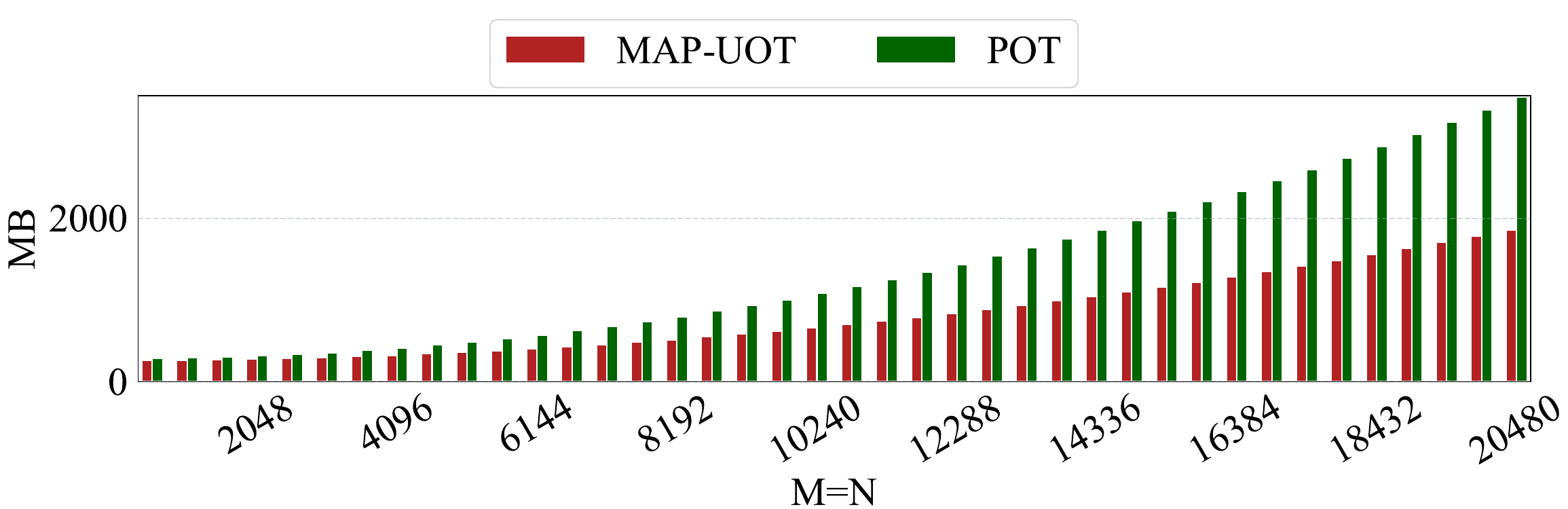}}
\caption{Peak memory consumption during computation for UOT.}
\label{fig:GPU_space}
\end{figure}

\begin{figure}
     \centering
     \begin{subfigure}[b]{0.22\textwidth}
         \centering
         \includegraphics[width=\textwidth]{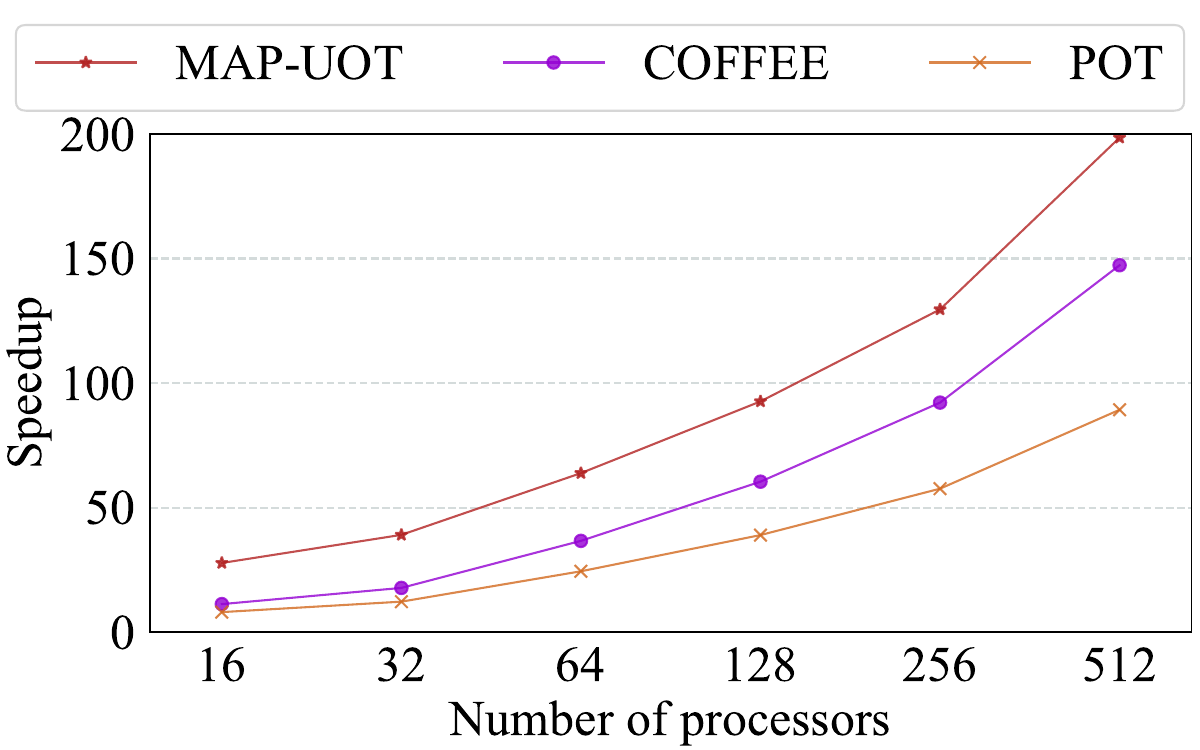}
         \caption{8 processors per node}
         \label{fig:Tianhe_speedup_0}
     \end{subfigure}
     \begin{subfigure}[b]{0.22\textwidth}
         \centering
         \includegraphics[width=\textwidth]{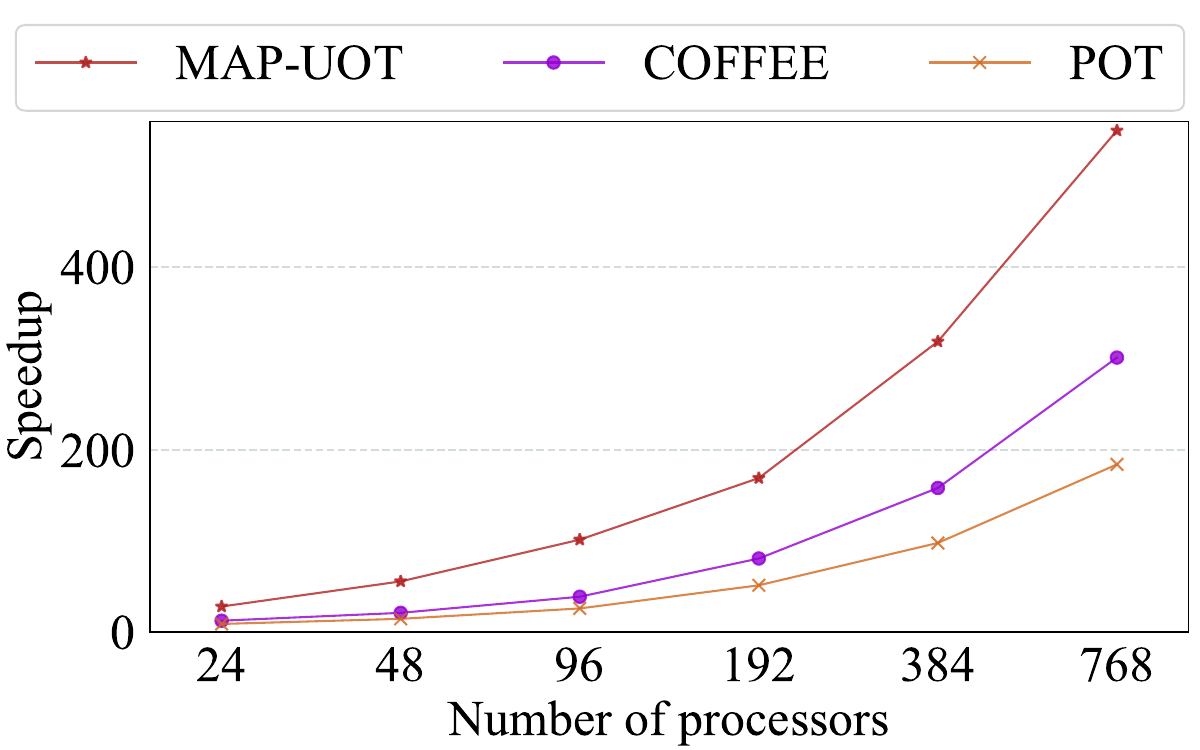}
         \caption{12 processors per node}
         \label{fig:Tianhe_speedup_1}
     \end{subfigure}
     \caption{Scalability of UOT on Tianhe-1 supercomputer.}
     \label{fig:Tianhe}
\end{figure}

\begin{figure}
     \centering
     \begin{subfigure}[b]{0.22\textwidth}
         \centering
         \includegraphics[width=\textwidth]{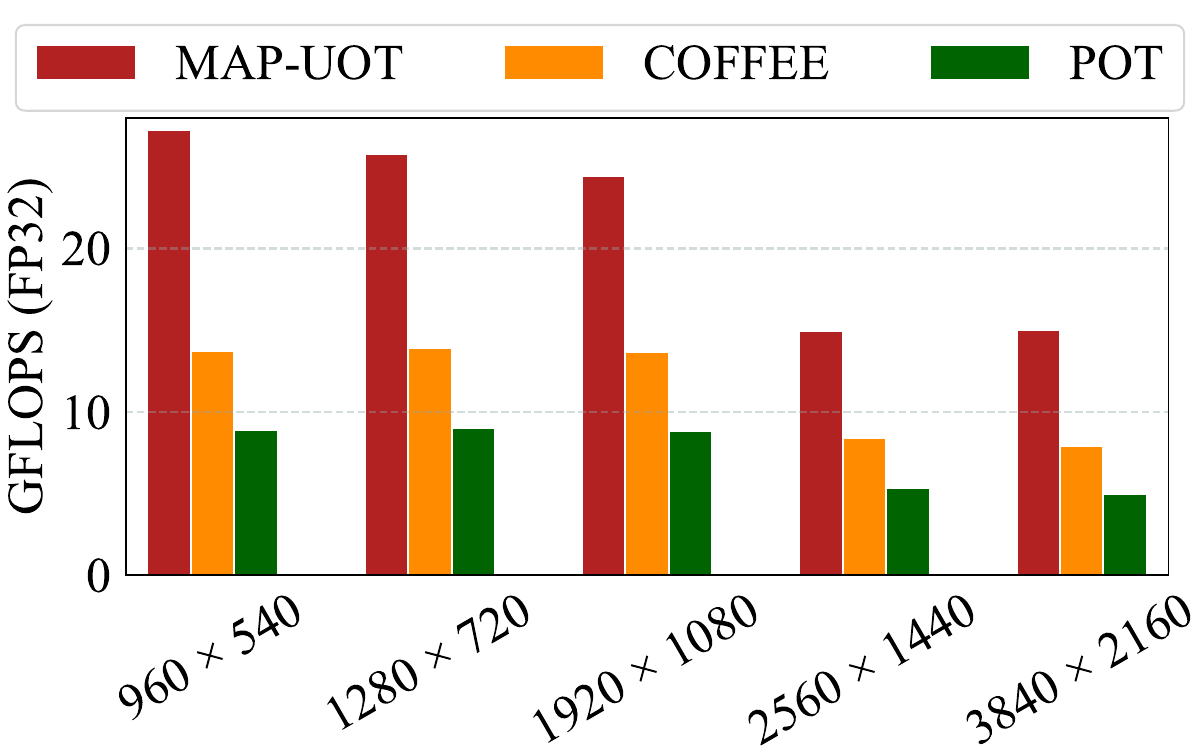}
         \caption{12900K}
         \label{fig:cpu_app}
     \end{subfigure}
     \begin{subfigure}[b]{0.22\textwidth}
         \centering
         \includegraphics[width=\textwidth]{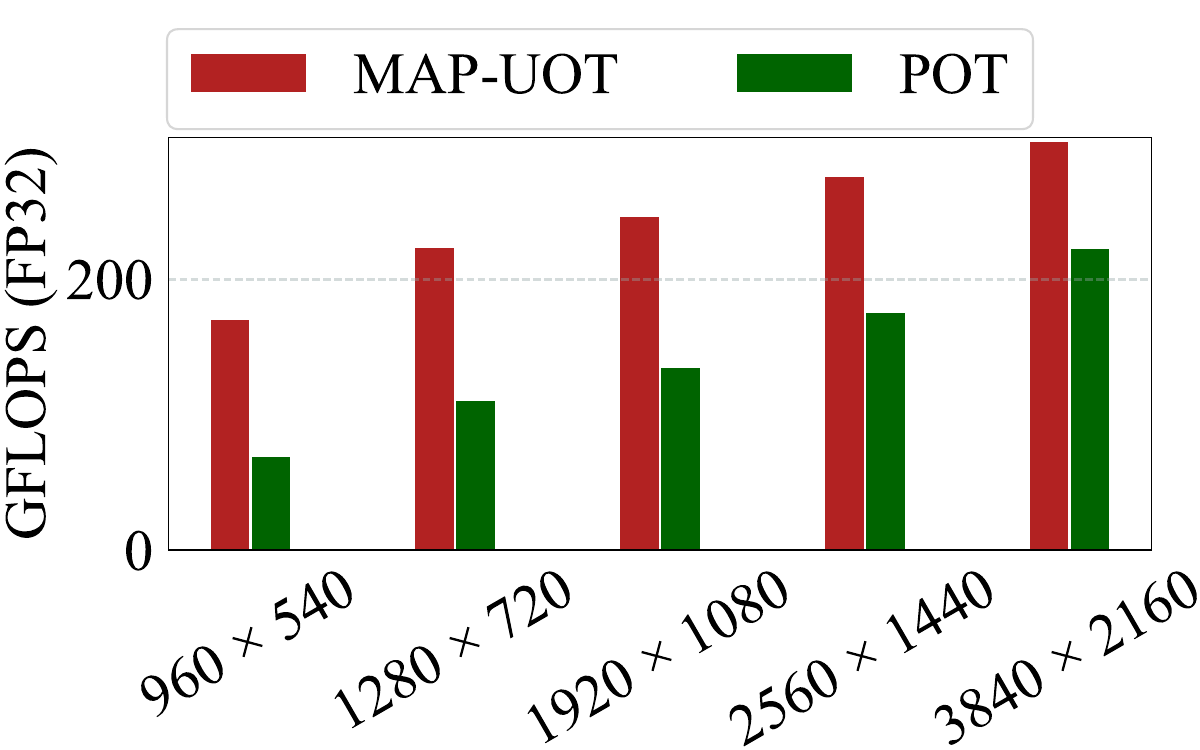}
         \caption{RTX 3090 Ti}
         \label{fig:gpu_app}
     \end{subfigure}
     \caption{Performance of UOT on the domain adaptation application.}
     \label{fig:app}
\end{figure}

\subsection{Performance of UOT on 12900K}

\subsubsection{Single-threaded Performance Improvement}In this experiment, we evaluate the single-threaded performance improvement MAP-UOT brings on 12900K. As shown in Figure~\ref{fig:CPU_new}, it is observed that MAP-UOT achieves significant and sustained performance improvements over others, both on square and rectangular matrices. For example, when $M = 1024, N = 10240$, the performance of MAP-UOT is 2.7X/2.2X over POT/COFFEE. It is found that the performance improvement gradually stabilizes, and the improvement is more significant with small matrix size. The reason is that the rows of the small matrix can be easily accommodated in the cache, which is more beneficial to our design. In general, MAP-UOT achieves up to 2.9X/2.4X with an average of 1.9X/1.6X performance improvement over POT/COFFEE.

\subsubsection{Scalability of Multiple Threads}In this experiment, we demonstrate the scalability of MAP-UOT over others on 12900K, and experimental results are normalized by the performance of single-threaded POT execution. As shown in Figure~\ref{fig:CPU_scalable}, MAP-UOT exhibits an average of 7.2X speedup with 16 threads, while POT/COFFEE exhibits an average of 3.3X/4.0X speedup in the same situation. As a result, MAP-UOT shows better scalability. It is found that parallel efficiency is reduced. Because the UOT algorithm is still memory-bound even though it has been significantly optimized. This results in threads not being able to release full computing performance as there are already enough memory requests to fully saturated the bandwidth, therefore leading to competition for bandwidth resources.

\subsubsection{Reduction of Cache Misses}In this experiment, we compare cache miss rate with and without MAP-UOT to show the impact of it on reducing cache misses. As shown in Figure~\ref{fig:CPU_cache}, it is observed that MAP-UOT significantly reduces cache misses over POT/COFFEE on all matrix sizes, both for L1 and L2 cache. For example, when $M = N = 4096$, L1 cache miss rate dropped by 57.4\%/39.4\%, L2 cache miss rate dropped by 79.2\%/64.3\% compared to POT/COFFEE. This is because MAP-UOT always traverses the matrix in row order and therefore access data in a cache-friendly way, resulting in improved performance.

\subsubsection{Impact of False Sharing}In this experiment, we demonstrate the false sharing problem in Pthreads to show MAP-UOT is not affected by this issue. It is noted that in shared memory parallel programs, there is a hidden problem causing performance degradation in general, that is, false sharing. False sharing occurs when multiple threads access multiple variables which are resided on the same cache line and need to modify at least one of them. The occurrence of false sharing will greatly reduce the performance of parallel algorithms. However, this is rarely the case with MAP-UOT. Because different threads are always accessing different rows of the matrix and different rows of $NextSum_{col}$. Besides, $Sum_{row}$ is a private variable of the threads, not affected by false sharing. Assuming a cache line size of 64B, as long as the number of columns in the matrix exceeds 16 and the data is memory aligned, the invisible problem of false sharing can be eliminated. As shown in Figure~\ref{fig:CPU_pthreads}, we demonstrate the L1 cache miss rate for some small, rectangular or large matrices under different threads. It is found that there is no significant increase in cache miss rate as the number of threads increases.

\subsection{Performance of UOT on RTX 3090Ti}

\subsubsection{Overall Performance Improvement}In this experiment, we evaluate the performance improvement MAP-UOT brings on RTX 3090 Ti. As shown in Figure~\ref{fig:GPU_new}, once again, it is observed that MAP-UOT achieves significant and sustained performance improvements over POT, both on square and rectangular matrices. For example, when $M = N = 4096$, the performance of MAP-UOT is 1.63X over POT. It is found that the performance improvement gradually stabilizes, and when the matrix size is small, both our and POT implementations degrade performance compared to when the matrix size is large. This is because when the matrix size is small, the full performance of the GPU cannot be released, since it will only run a few blocks, which makes some streaming multiprocessors in an unused state. For the case where fewer threads participate, the relative performance of the GPU will decrease. Overall, MAP-UOT achieves up to 3.5X with an average of 1.6X performance improvement over POT.

\subsubsection{Increment of Global Throughput}In this experiment, we compare global throughput with and without MAP-UOT to show the impact of our optimizations on RTX 3090 Ti using NVIDIA Nsight Compute toolkit (Ncu). As mentioned in Section~\ref{sec:motivation}, it is believed that this is one of the critical factors affecting the performance of UOT. As shown in Figure~\ref{fig:GPU_throughput}, it is observed that MAP-UOT significantly increases global throughput over POT on all matrix sizes, both for load and store throughput. For example, when $M = N = 4096$, global load throughput increased by 22.7\%, and global store throughput increased by 46.2\% compared to POT.

\subsubsection{Reduction of Space Consumption}In this experiment, we compare peak memory consumption over POT on RTX 3090 Ti with different matrix size. As shown in Figure~\ref{fig:GPU_space}, in all cases, MAP-UOT always uses less memory. For example, when $M = N = 4096$, MAP-UOT consumes 323 MB, which is 21.8\% less than POT.

\subsection{Scalability of UOT on Tianhe-1 Supercomputer}

In this experiment, we scale up MAP-UOT to run on Tianhe-1 supercomputer to verify the scalability of large-scale deployment. Instead of the Pthreads used in previous experiments, we use MPI to conduct this experiment because this is required for the multi-node distributed memory parallel programming. However, the idea is the same as the algorithm on a multi-core CPU. That is, we use standard MPI\_Allreduce~\cite{thakur2005optimization} function to replace Algorithm~\ref{alg:Pthreads_algorithm}, Line 16-20, aiming to calculate the sum of columns of the whole matrix. Then we use the mpi4py~\cite{smith2016performance} Python library to make the MAP-UOT achieve multi-node parallelism. Figure~\ref{fig:Tianhe} shows the scalability of MAP-UOT over POT and COFFEE on Tianhe-1 supercomputer. Here we choose different number of processors to process the same matrix of size $M = N = 20480$. We set two different cluster configurations, 8 processors per node and 12 processors per node, to demonstrate applicability. The results are normalized by the performance of single-threaded POT execution. The experimental results show that MAP-UOT not only outperforms POT and COFFEE but also exhibits the best scalability. The acceleration can reach up to 199X over 89X/147X of POT/COFFEE with 512 processors and 550X over 184X/301X of POT/COFFEE with 768 processors.

\subsection{Performance of UOT on Application}

In this experiment, we evaluate the performance improvement in the end-to-end application on 12900K and RTX 3090 Ti. We choose image color transfer~\cite{ferradans2014regularized} in domain adaptation to conduct the experiment, this application performs color normalization across several images and is widely used in computer vision and photography. As shown in Figure~\ref{fig:app}, MAP-UOT can have a consistently significant contribution to performance improvement on two architectures. For example, when matrix size is $1920 \times 1280$, the performance improvement of MAP-UOT is 2.77X/1.79X over POT/COFFEE on 12900K and 1.83X over POT on RTX 3090Ti. 

\section{Conclusion and Future Work}
\label{sec:conclusion}

In this work, we focus on the UOT algorithm and conduct in-depth analysis to discover the performance bottlenecks of UOT, and it is discovered that the UOT algorithm is heavily memory-bound.
We then propose MAP-UOT, a memory-efficient approach based on the discovered performance bottlenecks to optimise it.
A series of experiments confirm that MAP-UOT has a large improvement over other implementations on 12900K and RTX 3090 Ti.
Moreover, it shows good scalability on Tianhe-1 supercomputer.
In the future, we will combine our general design with specific architecture for further optimization, for example, taking advantage of complex structure of 12900K's performance cores and efficient cores.
Moreover, we will explore how to apply our approach to sparse matrices, since data stored in sparse matrices is common in the field of AI.
This will be another challenging work.

\bibliographystyle{abbrv}
\bibliography{scy}

\end{document}